# Panorama behaviors of Holographic Dark Energy models in $f(R,T)$ gravity

A.Y.Shaikh*, K.S.Wankhade

*Department of Mathematics, Indira Gandhi Mahavidyalaya , Ralegaon-445402.India.

Department of Mathematics, Y.C.Science Mahavidyalaya , Mangrulpir-444403.India.

**e-mail**:-shaikh_2324ay@yahoo.com

**Abstract**:

Classes of solutions of field equations in $f(R,T)$ gravity for a Bianchi sort I (Kasner form) space-time with matter and Holographic Dark Energy (HDE) is mentioned. Precise solutions of field equations are obtained with volumetrical power and exponential expansion laws .The physical and geometrical parameters of the models are mentioned well. The statefinder diagnostic pair and jerk parameter are analyzed to characterize utterly totally different phases of the universe.

# 1. Introduction

Observational information from the Cosmic Microwave Background (CMB) (Spergel et. al. (2003)), type IA Supernovae (SNe) ( Riess et. al.(1998) and Large Scale Structure (LSS)( Tegmark et.al.(2004)) indicates that our Universe is fast and increasing. The outre cosmic fluid having sturdy negative pressure makes the universe to accelerate its enlargement is that the Dark Energy (DE). The constant $\Lambda$ or LCDM model whose equation of state (EoS) parameter is w = -1 is that the easy candidate of the DE. The quintessence field models (Wetterich (1988),Ratra and Peebles(1988)),the phantom model(Caldwell(2002),Nojiri and Odintsov(2003)), k-essence(Chiba et.al(2000),Amendariz-Picon et al(2000,2001)), tachyon field (Sen(2002),Padmanabhan and Chaudhury(2002)) square measure the opposite candidates of dark energy models together with Chaplygin gas (Bento et.al(2002), Kamenchik et al.(2001)),Holographic Dark Energy ( Zhang et.al(2011),Granda and Oliveros (2009)) and then on. The modified theories of gravity square measure attracting presently many relativists to research DE models. Because of its ability to clarify many problems in cosmology and uranology, recently several cosmologists and astrophysicists have studied $f(R,T)$ theory. In this theory, the attractive force Lagrangian is given by associate impulsive operate of the Ricci scalar $R$ and of the trace of $T$ of the strain energy tensor. It's to be noted that the dependence of $T$ is also iatrogenic by exotic imperfect fluid or quantum effects. Katore and Shaikh (2012) investigated Kantowaski-Sachs cosmological models with anisotropic dark energy in $f(R,T)$ gravity. Reddy et al. (2013) have mentioned Bianchi type-III dark energy model in $f(R,T)$ gravity. Kaluza-Klein dark energy model square measure studied by Sahoo and Mishra (2014) within the presence of wet dark fluid in $f(R,T)$ gravity. Rao and Papa Rao (2015) have investigated five dimensional Kaluza-Klein reference frame within the presence of aeolotropic dark energy in $f(R,T)$ gravity. State finder diagnosing for holographic dark energy models in $f(R,T)$ gravity square measure studied by Singh and Kumar (2016). Shaikh A.Y.(2016) mentioned a self consistent system of Plane cruciform field of force and a binary mixture of perfect fluid and dark energy during a modified theory of gravity . Shaikh and Katore (2016) derived the precise solutions of the field equations with relevancy hypersurface-homogeneous Universe crammed with perfect fluid within the framework of $f(R,T)$ theory of gravity. Shaikh and Wankhade (2017) investigated Hypersurface-Homogeneous cosmological model in $f(R,T)$ theory of gravity with a term $\Lambda$. Moraes and Sahoo (2017) created a cosmological model from the only non-minimal matter–geometry coupling among the $f(R,T)$ gravity formalism. Analysis concerning compact stellar structures , hydrostatic equilibrium configurations of strange stars, aeolotropic stellar filaments evolving below enlargement free condition and also the dynamic stability of cutting viscous aeolotropic fluid were mentioned (Sharif and Siddiqa(2018),Yousaf et.al.(2018),Deb et. al.(2018a,2018b),Zubair et.al.(2018),Azam et.al.(2018)). Srivastava and Singh (2018) have obtained new holographic dark energy model with constant bulk consistence in $f(R,T)$ gravity . Moraes et.al.(2019) projected a brand new exponential form operate in hole pure mathematics among modified gravity. Moraes and Sahoo (2019) investigated wormholes in exponential $f(R,T)$ gravity. Recently, Reddy and Aditya (2019) have investigated Dynamics of perfect fluid cosmological model within the presence of scalar field in $f(R,T)$ gravity.

As a candidate of dark energy, HDE is associate rising model created by holographic principle (Guberina et al. 2007). A cosmological version of holographic principle was projected by Fischler and Susskind (1998) .Cohen et al. (1999), Horova and Minic (2000), Thomas (2002),

Hsu (2004), Li (2004), Setare (2007), Banerjee and Pavon (2007) and Sheykhi (2009) have investigated many aspects of holographic dark energy. Interacting changed HDE within the Kaluza-Klein universe is explored by Sharif and Jawad (2012). Samanta (2013) studied B-V reference frame HDE cosmological model among the presence of quintessence. Minimally interacting HDE square measure mentioned by Sarkar and Mahanta (2013), Sarkar (2014a, 2014b, 2014c) for aeolotropic models normally scientific theory. Jawad et al.(2015) have mentioned MHRDE in Chameleon with non-minimally matter coupling of the scalar field and its thermodynamic consequence. Reddy et.al(2016) have studied five dimensional spherically cruciform holographic dark energy cosmological model within the presence of minimally interacting fields; within the framework of Brans and Dicke (1961) theory of gravitation. Rao and Prasanthi (2017), Reddy (2017), Naidu et al. (2018) and Aditya and Reddy (2018) have explored MHRDE in scalar tensor theories of gravitations. Ricci dark energy model with bulk consistence has been investigated by Singh and Kumar (2018). Sharma and Pradhan (2019) analyzed the Tsallis Holographic Dark Energy (THDE) model mistreatment the statefinder diagnostic.

The main aim of this work is to check the aeolotropic Bianchi sort I (Kasner form) cosmological models within the presence of pressure less matter and modified holographic Ricci dark energy within the frame work of $f(R,T)$ gravity projected by Harko et al. (2011).

**2. Gravitational field equations of $f(R,T)$ gravity**

The $f(R,T)$ theory of gravity is the modification of General Relativity (GR). The field equations of $f(R,T)$ gravity obtained from the action

$$S = \frac{1}{16\pi G}\int \sqrt{-g}\, f(R,T) d^4 x + \int \sqrt{-g}\, L_m d^4 x, \tag{1}$$

where $f(R,T)$ is an arbitrary function of the Ricci scalar $(R)$ and trace of the stress energy tensor $(T)$ of the matter $T_{ij}$ $(T = g^{ij}T_{ij})$ and $L_m$ is the matter Lagrangian density.

The field equations in $f(R,T)$ theory of gravity for the function $f(R,T) = R + 2f(T)$ when the matter source is perfect fluid and given by (Harko et al. (2011))

$$R_{ij} - \frac{1}{2}Rg_{ij} = 8\pi T_{ij} + 2f'(T)T_{ij} + [2pf'(T) + f(T)]g_{ij}. \tag{2}$$

Here prime denotes differentiation with respect to the argument, $f(T)$ is an arbitrary function of the trace of stress energy tensor of matter and $p$ is the pressure of the matter source, which is a perfect fluid. For a detailed and systematic derivation of $f(R,T)$ gravity field equations one can refer to Harko et al. (2011).

## 3. Field equations of Bianchi Type I in Kasner form

Consider anisotropic Bianchi type I metric in Kasner form

$$ds^2 = dt^2 - t^{2q_1}dx^2 - t^{2q_2}dy^2 - t^{2q_3}dz^2, \tag{3}$$

where $q_1, q_2, q_3$ are three parameters that require to be constant. Let $S = q_1 + q_2 + q_3$, $\theta = q_1^2 + q_2^2 + q_3^3$, we get $R = (S^2 - 2S + \theta)t^{-2}$. The parameters $q_1, q_2, q_3$ will require to be constants and if at least two of the three are different, the space is anisotropic.

With the choice of the function $f(T)$ of the trace of the stress–energy tensor of the matter so that

$$f(T) = \mu T, \tag{4}$$

where μ is a constant ( Harko et. al. (2011)).

In this paper we assume that the universe is filled with matter and a hypothetical isotropic fluid as the holographic dark energy components. The energy momentum tensor for matter and holographic dark energy are defined as

$$T_{\mu\nu} = \rho_m u_\mu u_\nu\,; \quad \bar{T}_{\mu\nu} = (\rho_\lambda + p_\lambda)u_\mu u_\nu + g_{\mu\nu}p_\lambda, \qquad (5)$$

where $\rho_m$ is the energy densities of matter, $\rho_\lambda$ is the holographic dark energy and $p_\lambda$ is the pressure of the holographic dark energy.

Using comoving coordinates and equations (4) and (5), the $f(R,T)$ gravity field equations, (2), for metric (3) can be written as

$$\left[q_1(s-1) - \frac{1}{2}(s^2 - 2s + \theta)\right]t^{-2} = 8\pi p_\lambda - [-3p_\lambda + \rho_m + \rho_\lambda]\mu, \qquad (6)$$

$$\left[q_2(s-1) - \frac{1}{2}(s^2 - 2s + \theta)\right]t^{-2} = 8\pi p_\lambda - [-3p_\lambda + \rho_m + \rho_\lambda]\mu, \qquad (7)$$

$$\left[q_3(s-1) - \frac{1}{2}(s^2 - 2s + \theta)\right]t^{-2} = 8\pi p_\lambda - [-3p_\lambda + \rho_m + \rho_\lambda]\mu, \qquad (8)$$

$$\left[(s-\theta) - \frac{1}{2}(s^2 - 2s + \theta)\right]t^{-2} = -8\pi(\rho_m + \rho_\lambda) - [-p_\lambda + 3\rho_m + 3\rho_\lambda]\mu, \qquad (9)$$

where a dot here in after denotes ordinary differentiation with respect to cosmic time "$t$" only.

### 4. Isotropization and the solution

Define $R = \left(t^{q_1} t^{q_2} t^{q_3}\right)^{\frac{1}{3}}$ as the average scale factor so that the Hubble parameter in our anisotropic model may be defined as

$$H = \frac{\dot{R}}{R} = \frac{1}{3}\sum_{i=1}^{3} H_i \qquad (10)$$

where $R$ is the mean scale factor and $H_i = \frac{\dot{R}_i}{R_i}$ are directional Hubble's factors in the direction of $x^i$ respectively.

The anisotropy parameter of the expansion $\Delta$ is defined as

$$\Delta = \frac{1}{3}\sum_{i=1}^{3}\left(\frac{H_i - H}{H}\right)^2 \qquad (11)$$

in the $x, y, z$ directions, respectively.

The expression for scalar expansion and shear scalar is given by

$$\theta = 3H \qquad (12)$$

$$\sigma^2 = \frac{3}{2}\Delta H^2 \qquad (13)$$

The deceleration parameter is defined as

$$q = \frac{d}{dt}\left(\frac{1}{H}\right) - 1. \qquad (14)$$

The holographic dark energy density is given by

$$\rho_\lambda = 3(\alpha H^2 + \beta \dot{H}), \qquad (15)$$

with $M_p^{-2} = 8\pi G = 1.$

The continuity equation can be obtained as

$$\dot{\rho}_m + \dot{\rho}_\lambda + 3H(\rho_m + \rho_\lambda + p_\lambda) = 0. \qquad (16)$$

The continuity equation of the matter is

$$\dot{\rho}_m + 3H\rho_m = 0. \tag{17}$$

The continuity equation of the holographic dark energy is

$$\dot{\rho}_\lambda + 3H(\rho_\lambda + p_\lambda) = 0. \tag{18}$$

The barotropic equation of state is

$$p_\lambda = \omega_\lambda \rho_\lambda. \tag{19}$$

Using equations (15), (18) and (19), the EoS HDE parameter is obtained as

$$\omega_\lambda = -1 - \frac{2\alpha H \dot{H} + \beta \ddot{H}}{3H(\alpha H^2 + \beta \dot{H})}. \tag{20}$$

State finder parameters {r, s} are defined as (Sahni et al.2003)

$$r = \frac{\dddot{R}}{RH^3} \ , \ s = \frac{r-1}{3\left(q-\frac{1}{2}\right)}. \tag{21}$$

These parameters can be expressed in terms of Hubble parameter and its derivatives with respect to cosmic time as

$$r = 1 + \frac{3\dot{H}}{H^2} + \frac{\ddot{H}}{H^3} \ , \ s = \frac{-2(3H\dot{H} + \ddot{H})}{3H(3H^2 + 2\dot{H})}. \tag{22}$$

When $(r,s) = (1,1)$, we have cold dark matter (CDM) limit while $(r,s) = (1,0)$ gives $\Lambda$CDM limit. When $r < 1$ we have quintessence DE region and for $s > 0$ phantom DE regions.

In cosmology cosmic jerk parameter $j$, is used to describe models close to $\Lambda$CDM, which is a dimensionless quantity containing the third order derivative of the average scale factor with respect to the cosmic time. It is defined as (Chiba and Nakamura (1998))

$$j(t) = \frac{1}{H^3}\frac{\dddot{R}}{R} = q + 2q^2 - \frac{\dot{q}}{H}, \tag{23}$$

where $R$ is the cosmic scale factor, $H$ is the Hubble parameter and the dot denotes differentiation with respect to the cosmic time.

The stability or instability of the model depends upon the sign of $c_s^2 = \frac{dp_\lambda}{d\rho_\lambda}$, where $dp_\lambda$ and $d\rho_\lambda$ are cosmic time derivatives of pressure and density of dark energy, respectively. The models with $c_s^2 > 0$ are stable where as models with $c_s^2 < 0$ are unstable.

Using equations (6) and (7), we obtain

$$(q_1 - q_2)(s-1)t^{-2} = 0. \tag{24}$$

Equations (24) can be written as

$$\frac{d}{dt}\left(\frac{q_1}{t} - \frac{q_2}{t}\right) + \left(\frac{q_1}{t} - \frac{q_2}{t}\right)\frac{s}{t} = 0. \tag{25}$$

Let $V$ be a function of $t$ defined by

$$V = t^{(q_1+q_2+q_3)} = t^s. \tag{26}$$

Using equations (25) and (26), we get

$$\frac{d}{dt}\left(\frac{q_1}{t} - \frac{q_2}{t}\right) + \left(\frac{q_1}{t} - \frac{q_2}{t}\right)\frac{\dot{V}}{V} = 0. \tag{27}$$

Integrating the above equation, we get

$$\frac{q_1}{q_2} = d_1 \exp\left(x_1 \int \frac{1}{V} dt\right), \qquad (28)$$

where $d_1$ and $x_2$ are constants of integrations.

In view of $V = t^s$, we write $t^{q_1}, t^{q_2}, t^{q_3}$ in the explicit form

$$t^{q_1} = D_1 V^{\frac{1}{3}} \exp\left(X_1 \int \frac{1}{V} dt\right) \qquad (29)$$

$$t^{q_2} = D_2 V^{\frac{1}{3}} \exp\left(X_2 \int \frac{1}{V} dt\right), \qquad (30)$$

$$t^{q_3} = D_3 V^{\frac{1}{3}} \exp\left(X_3 \int \frac{1}{V} dt\right), \qquad (31)$$

where $D_i (i = 1,2,3)$ and $X_i (i = 1,2,3)$ satisfy the relation $D_1 D_2 D_3 = 1$ and $X_1 + X_2 + X_3 = 0$.

Since field equations (6)–(9) are four equations having five unknowns and are highly nonlinear, an extra condition is needed to solve the system completely. Here two different volumetric expansion laws are used, i.e.

$$V = at^b \qquad (32)$$

and

$$V = \alpha_1 e^{\beta_1 t}, \qquad (33)$$

where $a$, $b$, $\alpha_1, \beta_1$ are constants. The exponential law model exhibit accelerated volumetric expansion. In power law model, for $0 < b < 1$ the universe decelerates and for $b > 1$ the universe accelerates. In this way, all possible expansion histories, the power law expansion, (32), and the exponential expansion, (33) have been covered.

## 5. Model for power law

Using (32) in (29) -(31), the scale factors are obtained as follows:

$$t^{q_1} = D_1 a^{\frac{1}{3}} t^{\frac{b}{3}} \exp\left\{\frac{X_1}{a(1-b)} t^{1-b}\right\}, \tag{34}$$

$$t^{q_2} = D_2 a^{\frac{1}{3}} t^{\frac{b}{3}} \exp\left\{\frac{X_2}{a(1-b)} t^{1-b}\right\}, \tag{35}$$

$$t^{q_3} = D_3 a^{\frac{1}{3}} t^{\frac{b}{3}} \exp\left\{\frac{X_3}{a(1-b)} t^{1-b}\right\}, \tag{36}$$

where $D_i (i=1,2,3)$ and $X_i (i=1,2,3)$ satisfy the relation $D_1 D_2 D_3 = 1$ and $X_1 + X_2 + X_3 = 0$.

The mean Hubble's parameter, $H$, is given by

$$H = \frac{b}{3t}. \tag{37}$$

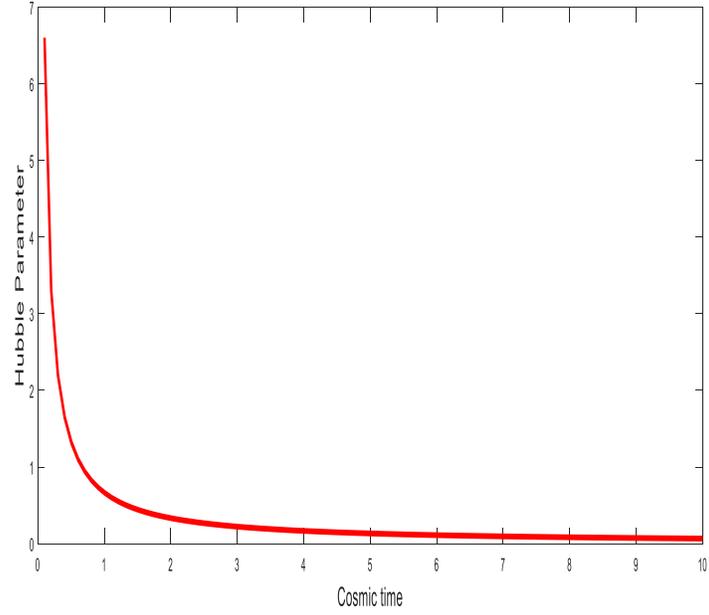

Figure 1. Variation of Hubble parameter against cosmic time for $b=2$.

The anisotropic parameter is given by

$$\Delta = \frac{3X^2}{a^2 b^2 t^{2(b-1)}}. \tag{38}$$

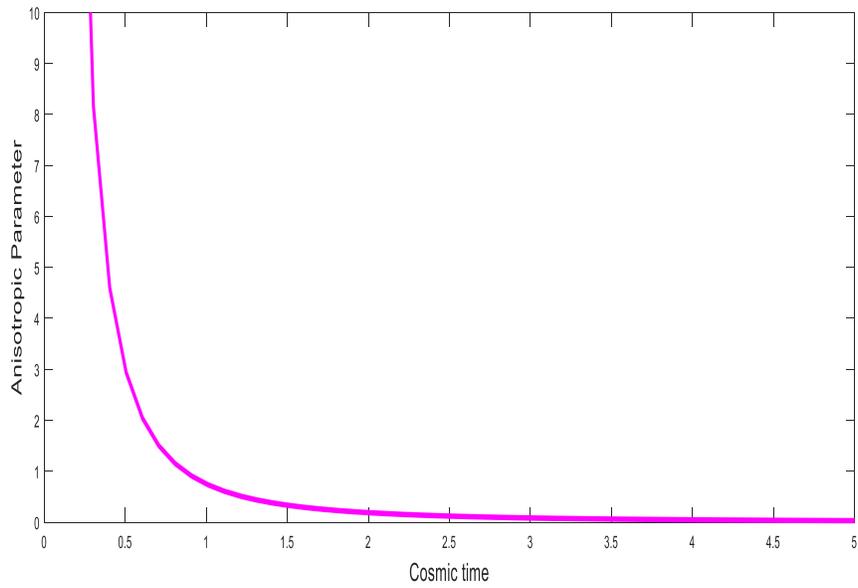

Figure 2. Variation of anisotropic parameter against cosmic time for $X=0.5, a=1, b=2$.

The dynamical scalars are given by

$$\theta = \frac{b}{t}. \tag{39}$$

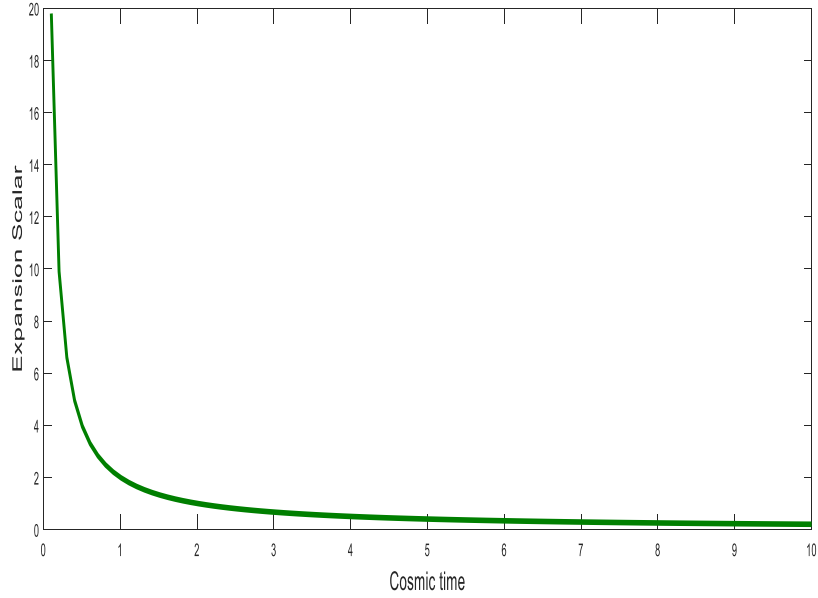

Figure 3. Variation of scalar expansion against cosmic time for $b=2$.

$$\sigma^2 = \frac{X^2}{2a^2 t^{2b}}, \tag{40}$$

where $X^2 = 2X_1^2 + X_2^2 = \text{constant}$.

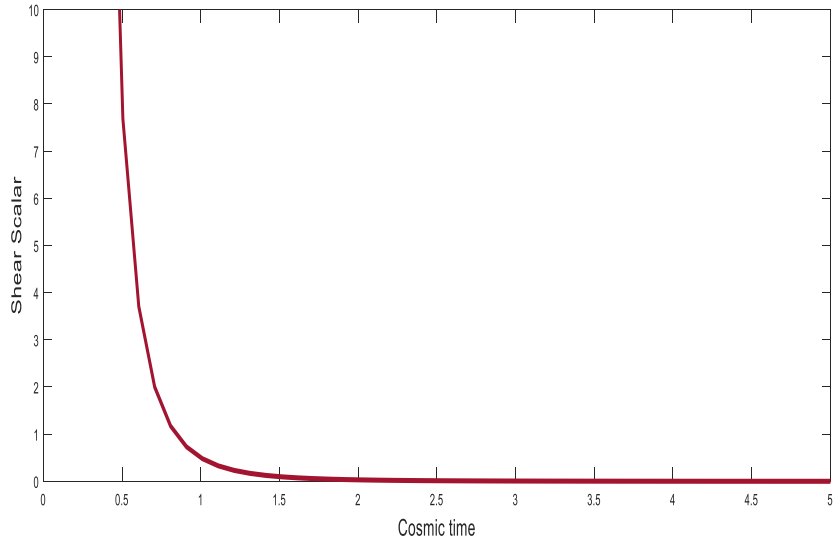

Figure 4. Variation of shear scalar against cosmic time for $X = 0.5, a = 1, b = 2$.

The deceleration parameter

$$q = \frac{3}{b} - 1. \tag{41}$$

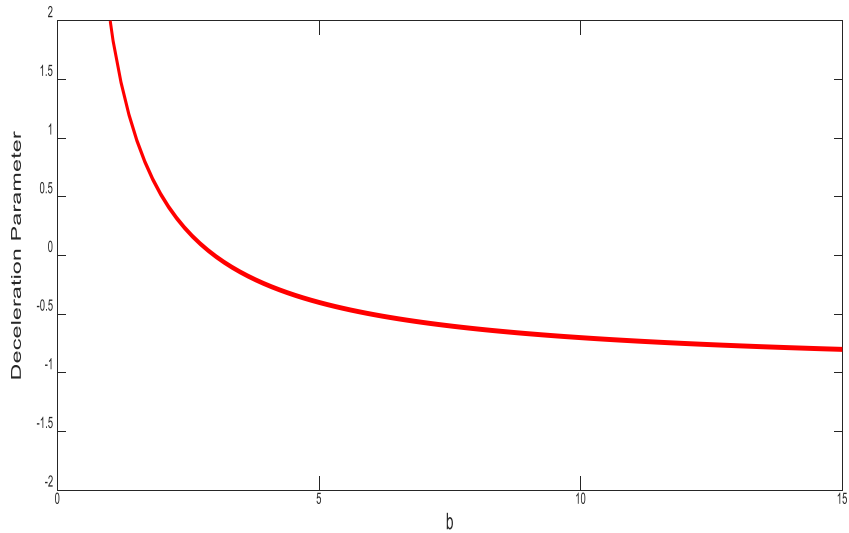

Figure 5. Variation of deceleration parameter against $b$.

The holographic dark energy density is obtained as

$$\rho_\lambda = \frac{b(\alpha b - 3\beta)}{3t^2}. \tag{42}$$

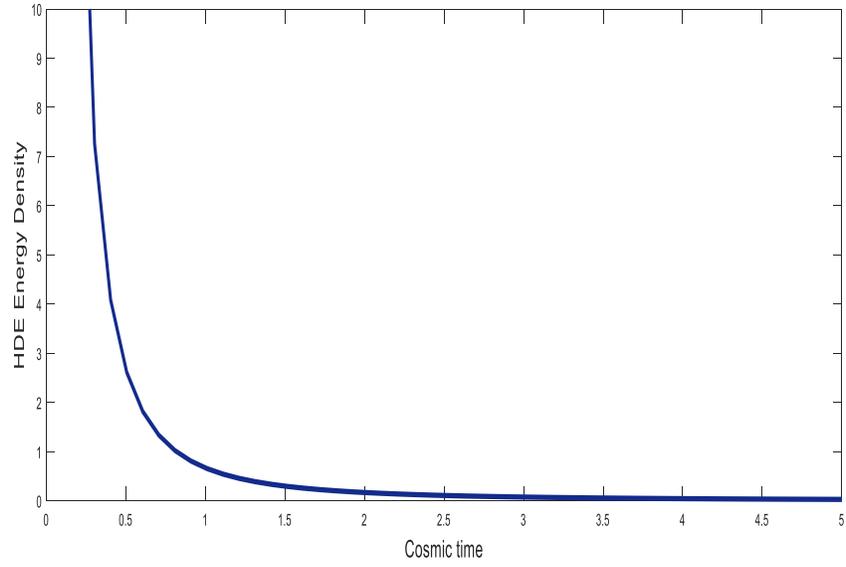

Figure 6. Variation of energy density of HDE against cosmic time for $, b=2, \alpha=2.5, \beta=1$.

The EoS parameter is given by

$$\omega_\lambda = \frac{2}{b} - 1. \tag{43}$$

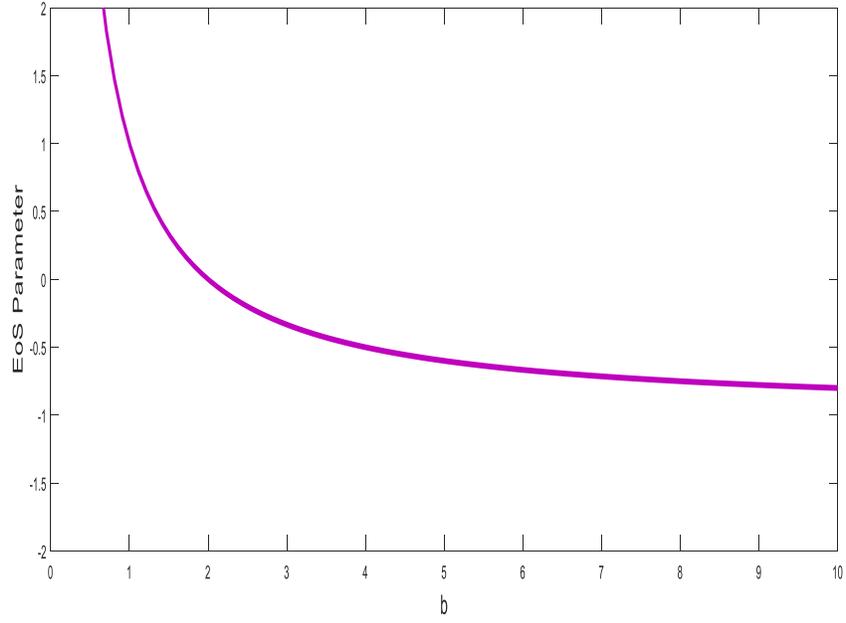

Figure 7. Variation of EoS parameter of HDE against $b$.

The holographic dark energy pressure is obtained as

$$p_\lambda = \frac{b(\alpha b - 3\beta)}{3t^2}\left(\frac{2}{b}-1\right). \tag{44}$$

The matter energy density

$$\rho_m = \frac{k_1}{at^b} \tag{45}$$

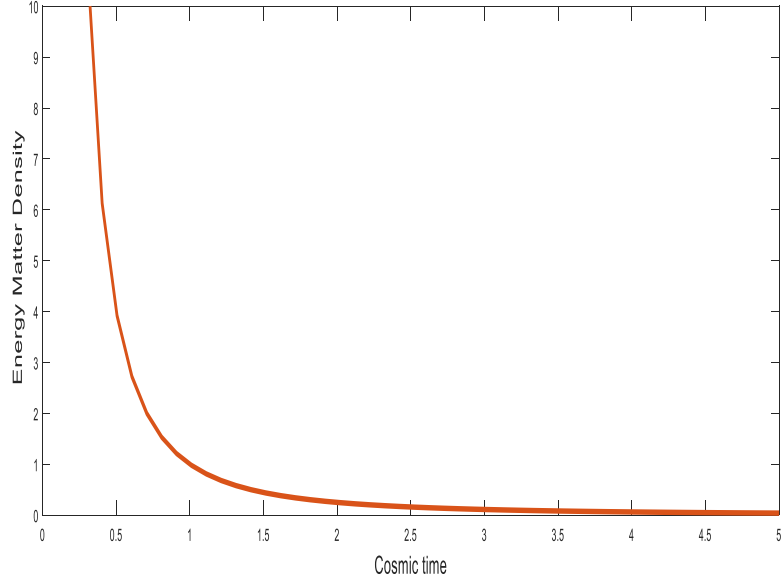

Figure 8. Variation of matter energy density against cosmic time for $k_1 = 1, a = 1, b = 2$.

The matter density parameter $\Omega_m$ and holographic dark energy density parameter $\Omega_\lambda$ are given by

$$\Omega_m = \frac{3k_1 t^{2-b}}{ab^2} \tag{46}$$

and

$$\Omega_\lambda = \frac{\alpha b - 3\beta}{b} \:. \tag{47}$$

Using equation (46) and (47), we get overall density parameter

$$\Omega = \Omega_m + \Omega_\lambda = \frac{3k_1 t^{2-b}}{ab^2} + \frac{\alpha b - 3\beta}{b} \tag{48}$$

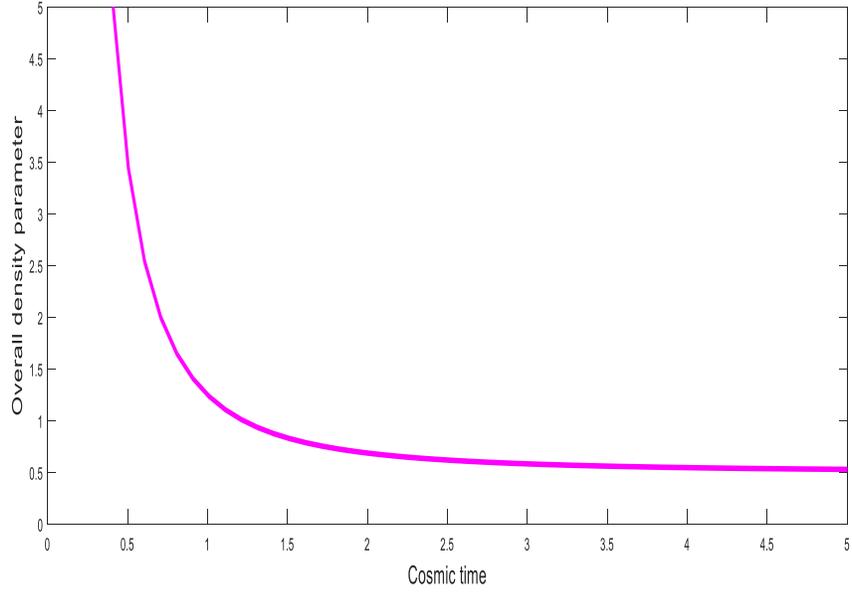

Figure 9. Variation of overall density parameter against cosmic time for $\alpha = 2.5, \beta = 1, a = 1, b = 1.2$.

The coincidence parameter

$$\bar{r} = \frac{\rho_m}{\rho_\lambda} = \frac{\dfrac{k_1}{at^b}}{\dfrac{b(\alpha b - 3\beta)}{3t^2}} \ . \tag{49}$$

The statefinder parameters are found as

$$r = \frac{b^2 - 9b + 18}{b^2} \quad \text{and} \quad s = \frac{2}{b} \tag{50}$$

i.e. $r = \dfrac{9s^2 - 9s + 2}{2}$ .

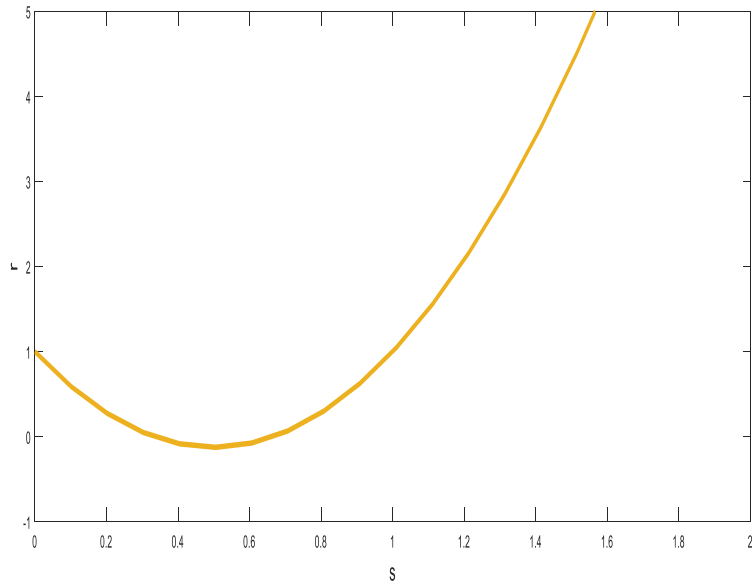

Figure 10. Variation of $r$ against $s$.

The jerk parameter yields

$$j(t) = \frac{b^2 - 9b + 18}{b^2} \qquad (51)$$

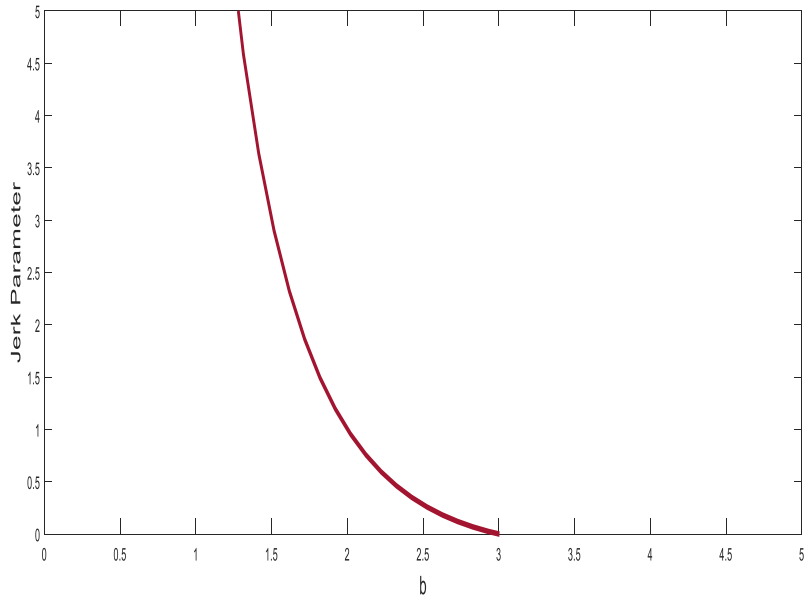

Figure 11. Variation of jerk parameter vs $b$.

From equations (42) and (44), it is observed that the ratio $c_s^2 = \dfrac{dp_\lambda}{d\rho_\lambda}$ is independent of cosmic time and totally depends upon the value of $b$.

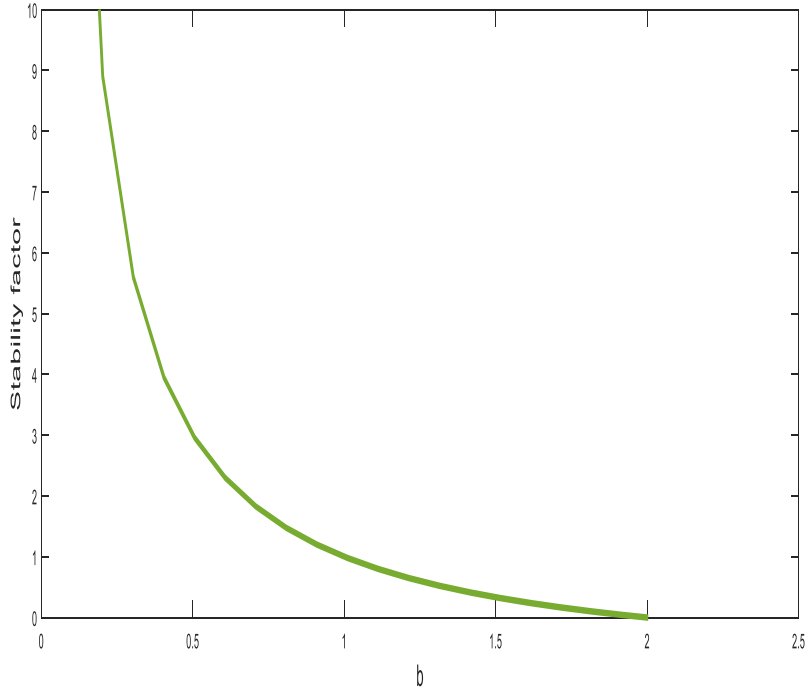

Figure 12. Variation of stability factor against $b$.

### 5.1. Physical behaviour of the power law model

We observe that the spatial volume $V$ is zero at $t=0$. So the model starts evolving with a big-bang kind singularity at $t=0$. It is seen that, near $t=0$, each the scalae factors vanish (Katore et.al(2011)). They increase with increasing time and reduces at massive time. So the model has associate in nursing initial singularity. At associate in nursing initial stage of growth, the Hubble parameter and shear scalar ar infinitely massive whereas with the growth of the universe each decreases (Katore and Shaikh(2012)). The graphical confirmation of shear scalar and Hubble parameter is shown in figures 1 and 4. Figure 2 depicts that the anisotropic parameter is additionally a perform of time , tends to infinite as $t \to 0$ and vanishes at $t \to \infty$ (Katore and Shaikh(2012)). From figure 3, we have a tendency to observe that once $t \to 0, \theta \to \infty$ and this means the inflationary situation at early stages of the universe(Katore and Shaikh (2014),Shaikh and Katore(2016)).

It is determined from the from figure 5 that the deceleration goes from positive to negative region freelance of time *t* and approaches the present value $q_0 = -0.725$ that matched with the determined value of Cunha et.al.(2009) . In recent times from the experimental knowledge , the constrain of the deceleration parameter is within the vary $-1 < q < 0$ (Perlmutter et.al.(1999),Riess et.al.(2004),Cunha et.al.(2009)).The negative value of the deceleration parameter represents the current acceleration of the universe. The deceleration parameter evolves from a positive region to a negative region showing a signature flipping.

Figure 7 depicts the variation of the Equation of State (EoS) parameter of HDE versus *b* . It's determined that Eos parameter of HDE may be a decreasing perform and converges to the negative value. By mere observation, it's clear that EoS parameter of HDE in early stages was positive (matter-dominated universe) and with the evolution of the Universe passes through phantom region $w < -1$ and approaches to $w = -1$ within the future, that is mathematically akin to the constant. Thus, the first matter dominated section soon regenerate to convert to DE phase.

Equation (44) shows that the pressure *p* of the universe is associate in nursing increasing perform of time *t*. At associate in nursing initial epoch, the HDE pressure begins from an outsized negative price and tends to zero at late time for $b > 2$ . The accelerated growth of the universe is thanks to the actual fact of Dark Energy i.e negative pressure. So the derived model is in smart agreement with the cosmological observations in accordance with the behavior of the pressure.

Figures 6 and 8 represent the behavior of energy density of HDE and matter energy density severally versus time $t$. It is seen from the graphs that HDE energy density and matter energy density decreases because the time will increase. The energy densities are positive throughout the evolution of the model. The energy densities are invariably positive and reduce with increasing time as determined within the model.

Figure 9 represents the behavior of overall density parameter versus time. With the cosmic evolution, the density parameter decreases with time. For correct alternative of the constants , the density parameter approaches to one that is in agreement with the experimental knowledge of the universe. The coincidence parameter is associate in nursing increasing perform of time $t$ . So the universe is dominated by HDE at early epoch of the universe and at later the universe is dominated by matter. This result's in smart accordance with the particular universe. The behaviour of the statefinder parameters ar displayed in figure 10.We see that the curve passes through the purpose . At some purpose of time our model corresponds to the ΛCDM model.

The jerk parameter is positive throughout the evolution of the universe as portrayed in figure 11. This shows that the universe exhibits a sleek transition of the universe from early speed to the current accelerated section that is in agreement with the current situation and observations of contemporary cosmology.

The stability issue is freelance of time . Figure 12 offers the graphical illustration of the satbility issue versus $b$ for the right alternative of constants. The model satisfies the difference $0 \leq C_s^2$ at late time evolution of the universe so the model don't admit superluminal fuctuations throughout late time (Vinutha et.al.(2019)).

## 6. Model for exponential law

Using (33) in (29) -(31), the scale factors are obtained as follows:

$$t^{q_1} = D_1 \alpha_1^{\frac{1}{3}} e^{\frac{\beta_1 t}{3}} \exp\left\{\frac{-X_1}{\alpha_1 \beta_1} e^{-\beta_1 t}\right\}, \tag{52}$$

$$t^{q_2} = D_2 \alpha_1^{\frac{1}{3}} e^{\frac{\beta_1 t}{3}} \exp\left\{\frac{-X_2}{\alpha_1 \beta_1} e^{-\beta_1 t}\right\}, \tag{53}$$

$$t^{q_3} = D_3 \alpha_1^{\frac{1}{3}} e^{\frac{\beta_1 t}{3}} \exp\left\{\frac{-X_3}{\alpha_1 \beta_1} e^{-\beta_1 t}\right\}, \tag{54}$$

where $D_i (i=1,2,3)$ and $X_i (i=1,2,3)$ satisfy the relation $D_1 D_2 D_3 = 1$ and $X_1 + X_2 + X_3 = 0$.

The mean Hubble parameter, $H$, is given by

$$H = \frac{\beta_1}{3}. \tag{55}$$

The anisotropy parameter of the expansion, $\Delta$, is

$$\Delta = \frac{3X^2 e^{-2\beta_1 t}}{\alpha_1^2 \beta_1^2}. \tag{56}$$

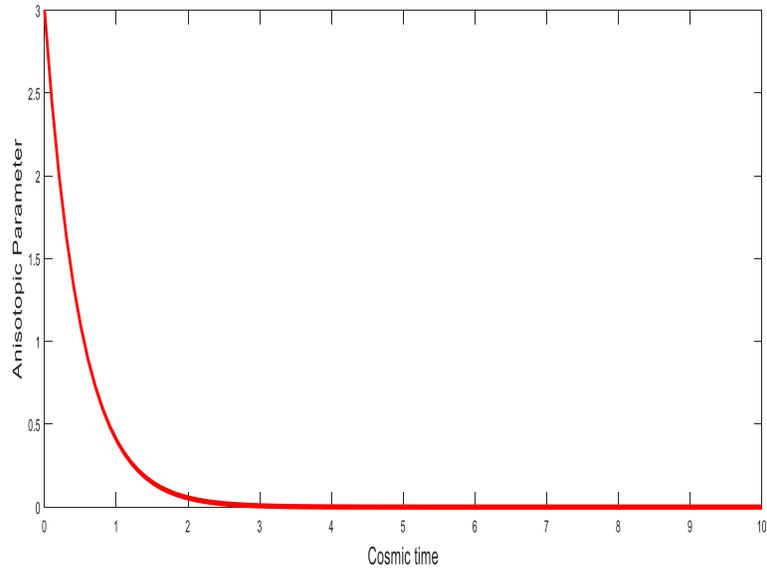

Figure 13. Variation of anisotropic parameter against cosmic time for $X = 0.5, \alpha_1 = 1, \beta = 2$.

The expansion scalar, $\theta$, is found as

$$\theta = \beta_1. \tag{57}$$

The shear scalar, $\sigma^2$, is found as

$$\sigma^2 = \frac{X^2 e^{-2\beta_1 t}}{2\alpha_1^2}, \tag{58}$$

where $X^2 = 2X_1^2 + X_2^2 = \text{constant}$.

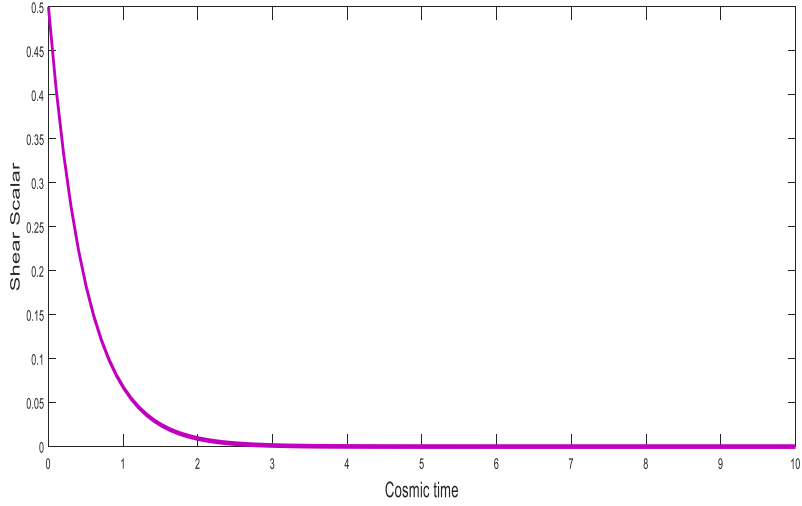

Figure 14. Variation of shear scalar against cosmic time for $X = 0.5, \alpha_1 = 1, \beta = 2$.

The deceleration parameter

$$q = -1. \tag{59}$$

The holographic dark energy density is obtained as

$$\rho_\lambda = \frac{\alpha_1 \beta_1^2}{3}. \tag{60}$$

The EoS parameter is given by

$$\omega_\lambda = -1. \tag{61}$$

The holographic dark energy pressure is obtained as

$$p_\lambda = -\frac{\alpha_1 \beta_1^2}{3}. \tag{62}$$

The matter energy density

$$\rho_m = \frac{k_2}{\alpha_1 e^{\beta_1 t}} \tag{63}$$

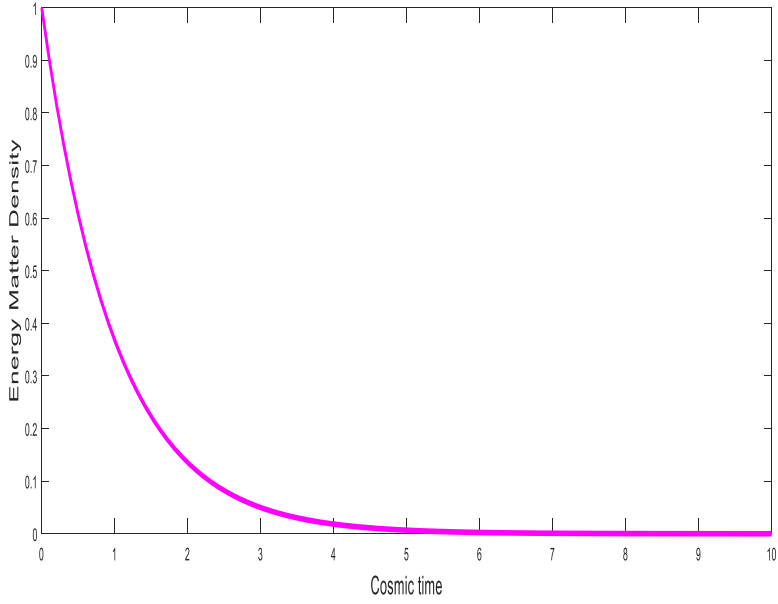

Figure 15. Variation of matter energy density against cosmic time for $k_2 = 0.5, \alpha_1 = 1, \beta = 2$.

The matter density parameter $\Omega_m$ and holographic dark energy density parameter $\Omega_\lambda$ are given by

$$\Omega_m = \frac{3k_2}{\alpha_1 \beta_1^2 e^{\beta_1 t}} \tag{64}$$

$$\Omega_\lambda = \alpha_1 \ . \tag{65}$$

Using equation (64) and (65), we get overall density parameter

$$\Omega = \Omega_m + \Omega_\lambda = \frac{3k_2}{\alpha_1 \beta_1^2 e^{\beta_1 t}} + \alpha_1 \tag{66}$$

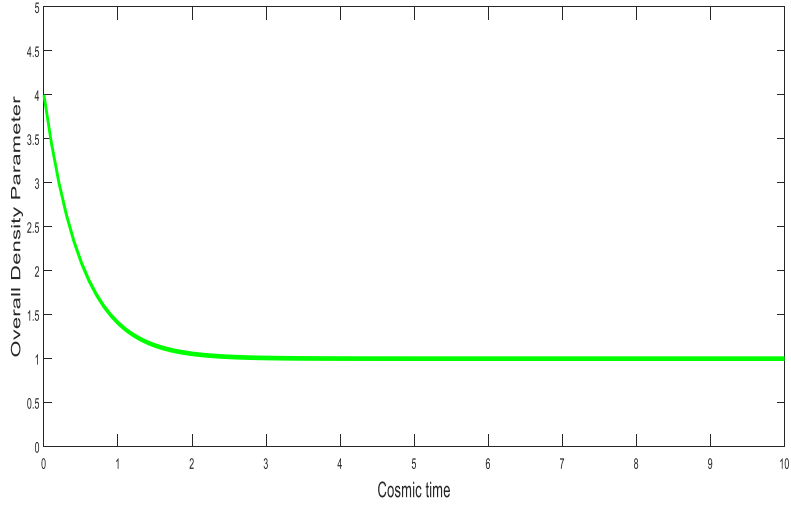

Figure 16. Variation of overall density parameter against cosmic time for $k_2 = 0.5, \alpha_1 = 1, \beta = 2$.

The coincidence parameter

$$\bar{r} = \frac{\rho_m}{\rho_\lambda} = \frac{\dfrac{k_2}{\alpha_1 e^{\beta_1 t}}}{\dfrac{\alpha_1 \beta_1^2}{3}} \quad . \tag{67}$$

The statefinder parameters are found as

$$r = 1 \text{ and } s = 0 \tag{68}$$

The jerk parameter yield

$$j(t) = 1 \tag{69}$$

The jerk parameter which is equal to 1 indicates a flat LCDM model.

### 6.1. Physical behaviour of the exponential expansion model

The abstraction volume is finite at $t=0$. It expands exponentially as $t$ will increase and becomes infinitely massive as $t \to \infty$. It is seen that, the metric potential admits constant values at the time $t=0$, afterward they evolve with time with none style of singularity and at last diverge to time. This is often in step with the big-bang state of affairs (Katore and Shaikh(2014),Katore et.al(2015),Katore and Shaikh(2015),Shaikh(2017)). The spinoff of mean Hubble parameter with reference to time vanishes implies the fast rate of growth of universe. So to explain the dynamics of the late time evolution, the derived model is thought of. At an initial epoch of the Universe, the mean anisotropic parameter of the growth is infinite and reduces with time and ultimately becomes zero as shown in figure 13. So property of the fluid doesn't support the property of growth. The growth scalar exhibits the constant worth that shows uniform exponential growth. The behavior shown in figure 14 specifies that the shear scalar is that the perform of time. It's infinite at an initial stage $(t=0)$, decreases with time, and vanishes for big worth of time $(t=\infty)$.

The universe decelerates for positive worth of deceleration parameter whereas it accelerates for negative one (Bennet et.al.(2003)). Equation (59) indicates that the universe is fast that is in step with this day observations that universe is undergoing the accelerated growth. Within the derived model we've got $\frac{dH}{dt} = 0 \Rightarrow q = -1$.

The physical behavior of holographic dark energy density is constant. It's discovered that HDE pressure is negative, that is that the reason for the accelerated growth of the universe. With applicable selections of integration constants and alternative physical parameters , the behavior of matter energy density is shown in fig. 15.It is discovered that the matter energy density decreases with time then tends to zero at $t \to \infty$. The overall density parameter approaches to one

, as portrayed in figure 16, that describes the flatness of universe and confirms this cosmological knowledge of the universe. It's clear that the coincidence parameter is that the decreasing perform of time. It's terribly massive at the first epoch of the universe however decreases monotonically at later time. Hence this universe is dominated by matter at early stages of the universe. For $r \to \infty, s \to -\infty$, the universe starts from a straight line Einstein static era and for $(r = 1, s = 0)$ goes to the LCDM model and a set newton's constant . The cold matter model containing no radiation is diagrammatic by for $(r = 1, s = 1)$. Equation (68) is comparable to the CDM cosmological model that the statefinder parameters square measure $\{r, s\} = \{1, 0\}$.

## 7. Conclusions

In this article, we've got studied the Bianchi type-I (Kasner form) universe in gravity in presence of matter and holographic dark energy. Here, we've got mentioned two models specifically law of power model and exponential law model. The observations obtained from these two models area unit conferred below:

**Power law volume enlargement model**

• The model starts evolving with a big-bang sort singularity at $t = 0$.

• The fastness goes from positive to negative region freelance of time t and approaches this worth $q_0 = -0.725$ .

• HDE energy density and matter energy density decreases because the time will increase.

• The jerk parameter is positive throughout the evolution of the universe.

**Exponential volume enlargement law model**

• The multiplier factor admits constant values at the time t=0, afterward they evolve with time with none varieties of singularity and at last diverge to time. This can be according to the big-bang state of affairs.

• q = -1 indicates that the universe is fast that is according to this day observations that universe is undergoing the accelerated (Katore et.al.(2011),Shaikh and Katore(2016)).

• It is discovered that the matter energy density decreases with time and so tends to zero at $t \to \infty$ (Shaikh et.al.(2019).

• The overall density parameter approaches to one that describes the flatness of universe and confirms this cosmological information of the universe.

This work can be extended to the other two classes of $f(R,T)$ gravity and also the other modified theories of gravity.